# Towards an Environmental Ethics of Artificial Intelligence

Nynke van Uffelen, Delft University of Technology, Faculty of Technology, Policy and Management, Delft, The Netherlands, n.vanuffelen@tudelft.nl, https://orcid.org/0000-0002-9500-8560

Lode Lauwaert, KU Leuven, Institute of Philosophy, Leuven, Belgium

Mark Coeckelbergh, University of Vienna, Department of Philosophy, Vienna, Austria

Olya Kudina, Delft University of Technology, Faculty of Technology, Policy and Management, Delft, The Netherlands

**Abstract**

In recent years, much research has been dedicated to uncovering the environmental impact of Artificial Intelligence (AI), showing that training and deploying AI systems require large amounts of energy and resources, and the outcomes of AI may lead to decisions and actions that may negatively impact the environment. This new knowledge raises new ethical questions, such as: When is it (un)justifiable to develop an AI system, and how to make design choices, considering its environmental impact? However, so far, the environmental impact of AI has largely escaped ethical scrutiny, as AI ethics tends to focus strongly on themes such as transparency, privacy, safety, responsibility, and bias. Considering the environmental impact of AI from an ethical perspective expands the scope of AI ethics beyond an anthropocentric focus towards including more-than-human actors such as animals and ecosystems. This paper explores the ethical implications of the environmental impact of AI for designing AI systems by drawing on environmental justice literature, in which three categories of justice are distinguished, referring to three elements that can be unjust: the distribution of benefits and burdens (distributive justice), decision-making procedures (procedural justice), and institutionalized social norms (justice as recognition). Based on these tenets of justice, we outline criteria for developing environmentally just AI systems, given their ecological impact.

## 1. Introduction

In recent years, there has been increasing attention to normative approaches to artificial intelligence (AI). More and more research is being conducted into normatively evaluating the preparation, development, and deployment of intelligent technologies. While it is possible to examine the facts about an AI system (such as what the system can or cannot do) without engaging in evaluation, a normative approach requires an adequate understanding of the facts of the technology, even though descriptive statements about AI systems are insufficient to ground normative claims.

Normative approaches focus strongly on transparency, privacy, safety, responsibility, bias, discrimination, and similar issues (Bakiner, 2023; van Wynsberghe, 2021). Although there is, of course, no doubt whatsoever that these topics deserve (a great deal of) attention, there are, however, reasons to believe that normative reflection on AI should not focus *exclusively* on these topics. Indeed, recently, an increasing number of academic papers have been dedicated to mapping the undesirable impacts of AI development and use on the environment, including ecosystems, nature, and animals (Falk & van Wynsberghe, 2023; van Wynsberghe, 2021).

With the concept of environment, we refer to the *natural* environment, including natural non-human entities such as ecosystems and animals. The environmental impact of AI refers, first, to the effects of AI infrastructure that facilitates training and deploying AI systems. The data centres in which AI models are trained and that process, for example, ChatGPT prompts, require significant amounts of electricity, which often, though not necessarily, leads to the emission of substantial amounts of greenhouse gases (Brevini, 2022; Crawford, 2021). Additionally, cooling these servers demands large quantities of water (Mytton, 2021; Zhao et al., 2022). This affects not only the lives of humans and (non-human) animals but also plants and ecosystems. Second, the environmental impact of AI refers to the impact of deploying AI systems in a specific context, such as in agriculture. As AI may generate recommendations, predictions, or actions, it may contribute to sustainability (this is called AI for sustainability or AI towards sustainability, see Falk & van Wynsberghe, 2023), yet it may also harm natural ecosystems and animals. For instance, AI-driven technologies such as drones and autonomous vehicles may frighten non-human animals, disrupt their habitats, or interfere with their migration patterns, all of which may (indirectly) impact humans as well.

Facts about the environmental impact of AI are normatively relevant, as they have undesirable consequences for the present and future lives of humans and other organisms. However, the environmental impact of AI does not necessarily imply that the development or use of AI is unjustified. In other words, undesirable environmental effects are insufficient for AI's development and use to be declared morally impermissible. Such a conclusion would lack nuance, as it is more reasonable to consider trade-offs, weighing the negative environmental impacts against the benefits of AI (alongside other potential drawbacks). A parallel can be drawn with motor cars: while their use predictably results in greenhouse gas emissions, this fact alone is insufficient to classify all car usage as unjustified. Similarly, in the case of AI, there is a need for criteria to distinguish between undesirable ecological impacts that are permissible and those that are impermissible. This raises a significant normative question: How to take into account the environmental impact of AI when assessing its moral (im)permissibility?

Although several authors call for ethical perspectives on the environmental impact of AI (e.g. Richie, 2022; van Wynsberghe, 2021), the topic remains largely unexplored in AI ethics literature in three regards. First, normative discussions about the justification of AI systems overwhelmingly focus on topics such as transparency, privacy, safety, responsibility, bias, and discrimination, largely ignoring the environmental impact of AI as an ethically relevant concern. Second, the environmental impact has consequences for humans, especially those living in places vulnerable to climate change or nearby datacentres and mines, but also for more-than-humans, such as animals, plants, and ecosystems. As such, the environmental impact of AI is problematic from an environmental ethics perspective (Moyano-Fernández & Rueda, 2024). Most papers in AI ethics cover issues such as transparency and discrimination, and thus its focus is mainly on what is good or just for people. In line with environmental ethics critiques on Western philosophy (Hourdequin, 2021), it can be argued that most AI ethics contributions are overly anthropocentric and thus the field should be reimagined and expanded in scope. Third, although numerous papers bring attention to facts about the environmental impact of AI systems, these works are generally descriptive in nature, as they report on and quantify the environmental impact of AI as an undesirable outcome, without discussing to what extent the environmental impact of specific AI systems are permissible or impermissible, and why (e.g. Bai et al., 2023; Zhou & Noonan, 2019). This gap in literature points to a need for AI ethics guidelines for developing and using AI models given their (predicted) environmental impact (e.g. Alloghani, 2023; Perucica & Andjelkovic, 2022).

The aim of this paper is not to provide a definitive answer to the question of on what grounds the undesirable environmental impact of an AI system would be justified or unjustified. Rather, the goal is to put the issue of the environmental impact of AI on the AI ethics research agenda, and to take initial steps toward filling this gap in the literature. We do so by presenting three concepts from environmental justice literature to ground the distinction between the permissibility and impermissibility of AI, given its ecological impact. We propose three sets of environmental justice principles that may be a starting point for formulating guidelines around evaluating AI systems, given their environmental impact. In doing so, we expand the scope of AI ethics towards considering the normative implications of the environmental impact of AI systems, which implies a less anthropocentric and more multispecies (Celermajer et al., 2021; Tschakert et al., 2021) approach to AI ethics.

We proceed as follows. Section two briefly describes some undesirable environmental impacts of AI development and use. Section three introduces three categories of environmental justice, which, as discussed in section four, can serve as a framework for evaluating AI systems as morally (im)permissible, given their (predicted) environmental impact. Section five outlines some objections and our responses. Section six concludes by summarizing our findings and proposing future directions.

**2. The environmental impact of AI**

Before zooming in on the negative environmental impacts of AI, we want to stress that many AI-enabled technologies have desirable environmental impacts and may help meet climate goals, such as electric cars and smart grid systems (e.g. Aldahmashi & Ma, 2022; Aurangzeb, 2019; Li et al., 2018). For example, in the Green Horizon project in Beijing (China), researchers use traffic cameras and weather stations, among others, to collect data on the distribution of particle pollution in the air. AI systems subsequently predict where and when pollution will occur, allowing the government to take targeted action to improve air quality. Consequently, particle pollution decreased by as much as 20 per cent.

The downside is that AI can be harmful to the environment. This pertains to various aspects of AI technologies throughout the entire life cycle. The environmental impact pertains, first, to the data gathering and training phase, the deployment of AI, the materiality and required resources of the data centres and hardware and resulting waste. Second, the environmental impact refers to the decisions made and actions taken based on the output that an AI model generates in specific contexts, such as energy infrastructure planning or agriculture.

First, the prevalent narrative of framing AI technologies as situated in 'the cloud' is erroneous, as it hides the fact that AI is not suspended in the air, but comes from and is embodied as materials. This hardware requires large quantities of metals and minerals, such as tin and silver. More concretely, about 35 per cent of the available amount of tin goes into making electronics; for silver, it is more or less 15 per cent (Crawford, 2021). Mining these metals and minerals is not without problems. Looking, for example, at pictures of Silver Peak, a mining site in the Nevada desert in the United States, one will see a large black spot next to the site. That is a pit filled with mud, brought to the surface when mining the metals and minerals needed to make computers. That mud contains toxic substances and is, therefore, potentially harmful to humans and other organisms (Crawford, 2021).

The hardware on which, among other things, large language models are trained is typically housed in what we call 'data centres'. Those big buildings are specialized facilities that house computer systems and associated components, such as storage and networking equipment. They ensure high performance, security, and reliability for storing and processing vast amounts of digital data. Worldwide, it is estimated that there are some 10,000 such centres (Minnix, 2024). That is, no small part of the earth's surface is occupied by matter for the purpose of training algorithms to suggest new bands or series to users, say. Although these facts may seem of little relevance at first glance, that is not the case. Indeed, where there are data centres, no organisms can live - no humans or other animals -, but also no trees or plants that can remove greenhouse gases from the air.

In addition, it is also important to consider what happens when AI systems are trained using hardware in data centres. Indeed, many computers working together in a datacentre emit much heat. To bring that temperature down, air conditioning is used, which requires large amounts of energy and water (Mytton, 2021). The more energy-intensive the computational processes in datacentres become (e.g. with the onset of Generative AI), the less one can rely on energy for air cooling as water presents new ways of cooling the datacentres (e.g. spraying the water onto the processors and into the air) and is cheaper compared to electricity (C. Li et al., 2018). To illustrate, an estimated five to nine per cent of total energy consumption is attributed to information technology, which includes AI (Caspart et al., 2022). Training a large AI system consumes about 1.3 gigawatt hours of electricity, as much power as more than 100 homes in the US annually (Vincent, 2024). According to research from Sweden, electricity demand from the AI sector is expected to be 15 times higher by 2030 than today (Brevini, 2022). If AI development continues at the same pace as in 2024, the energy required to run data centres will equal the current global energy consumption in 2040, regardless of AI development and use.

Consequently, albeit not necessarily, AI development and use is linked to the climate problem. Two per cent of global $CO_2$ emissions are currently said to be due to information technology and AI (Liu & Yin, 2024). Training a popular algorithm is responsible for emitting more than two thousand kilograms of CO2. By comparison, an average European flight emits about five hundred kilograms of carbon dioxide into the atmosphere per person, and the normal period of use of an average passenger car normally involves the emission of about forty thousand kilograms of $CO_2$ (Strubell et al., 2020). Some researchers estimate that by around 2040, 14% of global CO2 emissions will come from smart technologies (Belkhir & Elmeligi, 2018).

As we mentioned earlier, the 'environmental impact' not only refers to the effects of AI infrastructure for training and deploying algorithms but also to the effects of deploying AI systems in specific contexts, such as agriculture, on the natural environment. For example, automation may result in the loss of habitats. AI can be used in autonomous machines for harvesting crops, tilling soil, or managing forests, increasing the efficiency of these processes, yet it may also lead to the increased and accelerated use of land for agricultural or commercial purposes. Since many (non-human) animals rely on specific environments such as forests and grasslands for shelter, (non-human) animals lose their natural sources of safety when these areas are destroyed or disrupted by technologies like AI-driven agriculture or forestry (Karthik R & Ramamoorthy A, 2023). In addition, AI potentially has an impact on the food industry. For example, AI, including automated systems for breeding, tracking, and transporting animals, can optimise food processing, possibly increasing the demand for animal products and contributing to the neglect of animal rights and welfare (Todhunter et al., 2024).

**3. Concepts of environmental justice**

Knowledge about the environmental impact of AI raises new moral questions. This was already described by Mark Coeckelberg in 2021, who focused on the trade-off between freedom and AI technologies that influence human behaviour towards more sustainable practices, such as mobile applications that monitor and nudge household energy consumption (Coeckelbergh, 2021). Since not only the use of AI but also its training has a large environmental impact, several questions become urgent: given the environmental impact, when is it morally (un)justifiable and just to develop an AI system, and if it is permissible, how to make decisions about its design given the environmental impact?

The question of the moral permissibility, which includes the question of justice, of developing AI has hardly been asked in the literature. In 2023, Falk and Van Wynsberghe raised the following question in the context of AI systems that contribute to sustainability (AI for sustainability): "How useful can the impact of an AI system be towards sustainable ends if its own development and use defeats the purpose of its existence in the first place?" (Falk & van Wynsberghe, 2023, p.7). In other words, developing AI for sustainability might be fighting fire with fire. Moreover, in 2021, Van Wynsberghe stressed the importance of reflecting on the environmental impact in the innovation phase: "In a time when the world must commit itself to reducing carbon emissions, one has to ask if the emissions from algorithms that can play games (or do other menial tasks) is really worth the cost" (van Wynsberghe, 2021, p. 214). However, to our knowledge, no systematic responses have yet been formulated to the question of the moral permissibility of developing AI systems given their environmental impact.

To answer the normative question around AI and sustainability, we draw on environmental justice scholarship. *Environmental justice* emerged in citizen movements against climate and energy-related injustices (Holifield et al., 2017). Philosopher David Schlosberg analysed several citizen movements in the United States, concluding that their grievances can be understood using three concepts: distributive justice, procedural justice, and justice as recognition (Schlosberg 2004; 2007). Although several authors argue that the environmental impact of AI leads to environmental injustice (P. Li et al., 2024; Rakova & Dobbe, 2023; Taddeo et al., 2021), the link between EJ and the environmental AI has not yet been explored in the context of our research question. We propose to include Schlosberg's tripartite division to think about the environmental impact of AI.

First, *distributive justice* revolves around distributing goods, services, burdens and benefits in a society. Many modern political philosophers, such as John Rawls, regard this category as a core concern of justice. According to Rawls, the object of distributive justice is the distribution of 'primary social goods' in society, including rights and freedoms, power and opportunity, income and wealth, and self-respect as social bases (Rawls, 1971, p. 79). Although Michael Walzer argued that goods like leisure, education, divine grace, friendship and love can also be the object of distributive justice (Walzer, 1983), other philosophers think this goes too far and argue that distributive justice is only about distributing goods that are scarce and tangible, such as money, housing, and energy (Young, 1990).

Besides discussions about the *object* of distributive justice, there are also different *conceptions* of distributive justice. Although some argue that distributive justice should be about equal distribution of resources or of other goods, most Western philosophers justify some level of inequality, provided that certain conditions are met. Distributive principles may be based on specific interpretations and definitions of needs, wants or well-being, such as Amartya Sen and Martha Nussbaum's capabilities theories, (Nussbaum, 2011; Sen, 2009), merit, strict equality, equity, human rights, or dignity - and each principle has a different idea of when and to what extent inequalities are justified (Miller, 2017). Rawls is known for the so-called *difference principle*, in which differences are justified only if the position of the least advantaged is maximized (Rawls, 1999). Another theory is *sufficitarianism*, which portrays a just society as one in which everyone has "enough" or a sufficient level of resources, capabilities, or well-being, rather than striving for complete equality or maximizing overall welfare. In contrast to *egalitarianism* (which seeks equality) or *utilitarianism* (which aims to maximize a certain utility, such as happiness), sufficitarianism emphasizes ensuring that all individuals meet a basic threshold, and beyond this point, the relative distribution of goods is seen as less morally important. For example, Harry Frankfurt has famously argued that "With respect to the distribution of economic assets, what is important from the point of view of morality is not that everyone should have the same but that each should have enough" (Frankfurt, 1987, p. 21). Crisp has argued for a 'sufficiency' principle based on compassion (Crisp, 2003). There are also different views about where the threshold should be, in other words, about what constitutes 'enough'. Lastly, there are disagreements about the extent to which we should include future generations in balancing benefits and burdens; whether animals and ecosystems are only of instrumental importance, or should also be the subject of justice (Celermajer et al., 2021; Hourdequin, 2021; Tschakert et al., 2021); and what the appropriate scale of justice is in a specific context is, in other words, whether redistribution should be should be considered globally, or whether it is justifiable to limit the moral circle to a specific country or region (van Uffelen et al., 2024).

The second category of justice concerns *procedural justice*, which focuses on decision-making procedures: how and by whom are decisions made, and are these procedures fair? The distinction between distributive and procedural justice was already made by Rawls (Rawls, 1999). Justice, he argued, is not compatible with an authoritarian figure who distributes societal joys and burdens top-down; justice also implies a just decision-making procedure (Rawls, 1999, p. 118). Arguments for democratising decision-making around AI relate to the quality of outcomes, legitimacy, and intrinsic and relational reasons (Wielinga & Buijsman, 2024). However, there is also uncertainty about what constitutes a just decision-making procedure. There are various versions of democratic decision-making, in which stakeholders have more or less influence, and the criteria for decision-making can vary from a lottery to majority rule or consensus. Moreover, it remains unclear to what extent non-humans, future generations, and affected stakeholders beyond (supra)national borders should have a say and, if so, how and how significant their voices should weigh.

A third category refers to *justice as recognition*. This view represents a critique of the traditional focus on distributive justice in political philosophy (Young, 1990) and related views of human beings as atomistic and autonomous individuals (Anderson & Honneth, 2009). According to recent political philosophers such as Iris Marion Young, Nancy Fraser, Charles Taylor and Axel Honneth, justice is not merely a virtue of decision-making procedures or the distributive consequences of institutions for individuals in a society. Institutions, including the market, education, politics, and (AI) technologies, result from values and social relations. Therefore, not only their distributions and decision-making procedures but also the norms, values and social relations that (implicitly) constitute institutions and technologies should be critically scrutinised. (Mis)recognition is crucial to consider in relation to environmental justice, as climate policies and energy infrastructures and systems typically reflect the dominant status order and may thus be racist, speciest, or biased against the interests of the minority world, disrespecting and devaluing indigenous ways of life, the flourishing of nature, and the well-being of humans and more-than-humans (Álvarez & Coolsaet, 2020; Coolsaet & Néron, 2020; Holifield, 2012; Schlosberg, 2007).

Axel Honneth's and Nancy Fraser's theories of justice as recognition help us understand this tenet of justice. Fraser proposes the concept of *status order*, referring to cultural value hierarchies that can be institutionalised, for example in relation to gender, race, religion, age, or species (Fraser, 2000; Fraser & Honneth, 2003). Honneth, on the other hand, focuses on social relations, which he conceives as relations of recognition of love, law, and esteem, which find their expression in society's institutions (Honneth, 1995). The dominant status order and relations of recognition are institutionalised; however, when an unjust status order is institutionalised, there is misrecognition. Examples of misrecognition are sexism, racism, ageism, ableism, or speciesism. In such cases, certain identities or properties are seen as inferior ('less worthy'), or associated with negative traits, giving them a lower social status, that is, a lower place in the hierarchy of social values in the status order. In other words, misrecognition refers to institutional, cultural, and symbolic injustices. Justice as recognition thus expands the object of (in)justice: in addition to distributions and procedures, institutionalised norms, values and relations of recognition can also be unjust.

The three concepts of distributive, procedural, and recognition justice refer to three elements that can be (un)just about AI systems, namely its distributions of burdens and benefits, decision-making procedures, and the status order and relations of (mis)recognition embedded in the (design of the) technology. These three elements can or should be kept in mind when exploring the moral permissibility of developing an AI system, given its environmental impact.

**4. Towards an ethical framework for developing AI given its environmental impact**

When developing AI, it is important to reflect on the distribution of burdens and benefits, the decision-making procedure, and the cultural norms and relations of (mis)recognition institutionalised in the design. This reflection is one of applied political philosophy: under what conditions is developing an AI system fair, given its environmental impact? In short, it is crucial to formulate principles of justice in this context, which should be integrated into existing ethical guidelines around developing AI systems.

*4.1 Distributive justice*

We argue that ethical evaluations of AI systems should explicitly include the distributive consequences of the system, including its environmental impact. The benefits of AI systems are, for example, financial in nature or consist of more autonomy for a certain group of people, efficiency of a process or task, or well-being or pleasure for a group of users. The burdens of AI range from a loss of privacy or autonomy and harm due to bias to the disadvantages of data centres, including the aesthetic and cultural impact on the landscape and the environmental impact of the required electricity, materials, water, and related waste. For distributive justice reasons, it is essential to determine which actors, from what geographical areas, will benefit from AI, and where the burdens, including the negative ecological impacts of AI, will manifest.

One example in which the tension between distributed burdens and benefits plays out can be found in the context of smart grid systems, which are technologies that help stabilise the electricity grid. Because solar and wind power are less stable than fossil fuels, grid congestion can occur, sometimes leaving too little energy to meet our needs. Therefore, various smart grid technologies are being developed that do weather forecasts, interact with energy storage, and stabilise the grid (Aldahmashi & Ma, 2022; Aurangzeb, 2019; C. Li et al., 2018). Thus, AI helps manage energy supply and demand. The existence of this AI is justified because it contributes to a stable and more efficient national electricity grid, and thus this application of AI can be seen as contributing towards sustainability (Falk & van Wynsberghe, 2023). Smart grid systems mainly benefit countries that use it, such as the Netherlands and Belgium. On the other hand, AI in smart grids itself has a footprint due to its material, water, and electricity use, and as such, it also contributes to global warming, and the most affected people and regions (MAPAs) of climate change, most of which are in the Global South, bear little responsibility for the problem. It is crucial to reflect on the equity of this distribution.

It is thus crucial to formulate principles of distributive justice on a conditional basis, such as 'Developing an AI is fair if the burdens and benefits...'. Such concrete formulations can then underlie political guidelines. To our knowledge, such an attempt has not yet been made. Falk and Van Wynsberghe have attempted to formulate an answer to the question of the permissibility of AI given its ecological impact: "AI for Sustainability applications should only considered as such when they satisfy the condition that they also take into account the environmental damages and mitigate said damages through reduction of energy consumption" (Falk & van Wynsberghe, 2023, p. 8). This should consider "the greenhouse gas emissions generated by AI research against the energy and resource efficiency gains that AI can offer" (Falk & van Wynsberghe, 2023, p. 8). However, this answer is insufficient to shape guidelines. Equity is not the same as adding and subtracting advantages and disadvantages: it also matters which actors bear the benefits and burdens. For example, AI can have a large local impact on communities, especially in areas where water is scarce (P. Li et al., 2024), while people who benefit from its use are usually in areas remote from these AI production sites.

One possible formulation of principles of distributive justice in the context of the environmental impact of AI is this:

*Developing an AI is just if the environmental impact does not outweigh the benefits, that is, ...*
   *(a) if the benefits of the AI outweigh its environmental impact, and*
   *(b) if it maximises the benefits for the least well-off (Maximin), and*

*(c) if it promotes, or at least does not violate, a minimum level of well-being.*

A few comments are necessary. First, we propose to describe the benefits of AI in terms of capabilities, the UN SDGs (the United Nations Sustainable Development Goals), or a similar proposal that tries to capture human and animal welfare, to exclude AI systems that do not contribute to welfare, deployment, or sustainability. Developing systems that altogether do not contribute to well-being are not justified, in our view, given their environmental impact. Second, criterion (a) relies on a consequentialist conception of ethics, weighing likes and dislikes. A problem with criterion (a), however, is the incommensurability of values. It remains an open question how environmental impact should be weighed against ChatGPT's pleasure to hobbyists, or the time savings it can save companies. Moreover, we recognise the incompatibility of consequentialism with many intuitions of justice (Rawls 1971). Because of these problems, criterion (b) is included.

Criterion (b) emphasises the importance of social groups when it comes to justice. Injustice is often characterised by the fact that the pleasures belong mainly to one group (or country or species), while the burdens fall elsewhere (Young, 1990). AI systems may bring pleasure, efficiency, and financial gains for a group of privileged people, while the burdens worsen the situation of the globally least well-off, and this is generally considered an unjust situation. Criterion (b) thus allows for concerns around intergenerational justice, global justice and interspecies justice, as these actors are at risk of bearing most of the burdens of AI. Criterion (b) also acknowledges that an unequal distribution of burdens and benefits is not intrinsically problematic. Distributive inequality can be justified if the likes benefit the least well-off in society, while the strongest shoulders bear the heaviest burdens (Rawls, 1999). An AI can be emancipatory and promote social justice precisely by having an unequal impact on different social actors.

The third criterion (c) considers the importance of minimum thresholds that should be reached. This criterion assumes that a minimum level of well-being and social security should be guaranteed for everyone, including people in the Global South who are in vulnerable positions regarding the consequences of climate change. A sufficitarian approach to climate change could focus on ensuring that all people have access to the environmental resources necessary for a minimally acceptable quality of life and well-being (e.g., clean air, access to water, and protection from extreme weather events). Climate policies would prioritize helping those most vulnerable and affected by climate-related harms—ensuring that no one falls below the threshold of sufficient environmental well-being—before addressing concerns about inequality in carbon emissions or wealth. One could also require that everyone should have enough resources to lead a *sustainable* life or demand that *future generations* have access to a minimum threshold of resources. One could also argue that people's human rights should be respected at all times, from a deontological standpoint.

Although most principles for distributive justice formulated in Western philosophy are mainly anthropocentric, we want to stress that it is crucial to take into account the distributive impact of animals and nature as well as the impact on humans. However, much work in this area is yet to be done (Hourdequin, 2021). For example, what does Rawls's difference principle or the sufficitarian principle imply for non-humans and the environment: does the difference principle apply to the least advantaged species as well, ensuring that one should prioritize maximizing the position of endangered animals and ecosystems? How should we conceive animal and ecosystem well-being and capabilities, and what implications would this have for AI development? And can non-humans, such as rivers and landscapes, have rights (as is already the case in some places, e.g. Bellina, 2024), or would this entail consequences currently perceived as too radical? Such questions on non-anthropocentric distributive justice are only beginning to emerge, and much work is still to be done.

We present these three conditions as a starting point for discussion, and not as an end point or final conclusion. The aim of this paper is to provide a first iteration of criteria for the moral permissibility of AI development that incorporate the environmental impact, which can be ultimately incorporated in developmental guidelines, as a starting point or threshold that needs to be met before proceeding with the project. As such, the principles formulated in this paper can be considered as outlines or first steps to kick-start the discussion.

*4.2 Procedural justice*

A second tenet is just decision-making procedures for developing AI, given its environmental impact. In the context of AI, procedural justice translates as the following question: Who gets to decide whether or not and how this AI should be developed, and how should these decisions be made?

Currently, AI systems, such as ChatGPT, are developed by private companies. Decision-making power, therefore, lies solely with designers and investors. In this, they are only restricted by the guidelines issued by (supra)national governments. As long as companies stay within the legal frameworks, they can develop, train, and deploy AI systems as they please. In short, private actors can freely develop AI systems with large footprints. However, from an ethical perspective, this is questionable. The impact of AI systems, including the environmental impact, goes far beyond the actors making the decisions. The people who bear the burdens of its impact are usually not involved in the decision-making process. In addition, the voices of ecosystems, animals, and future generations are missing, or can only be indirectly and by proxy represented in decision-making.

Principles of procedural justice should be developed in this context in the form of 'The decision procedure to develop an AI is fair if...'. We make this proposal when it comes to procedural justice:

*Developing an AI is equitable if the decision-making procedure…*
  *(d) gives all stakeholders a voice, and*
  *(e) the voices of the most affected groups weigh significantly on the decision outcomes*

Criterion (d) reflects the all-affected principle: everyone affected by a decision, or every stakeholder, should have a say. AI's environmental impact affects not only humans, but also other species and future generations, and so they should also have a voice (Latour, 2017). 'Having a voice' here implies participation or involvement in decisions in the broadest sense: participation can be direct, indirect, or hypothetical. Decisions about AI systems often require technical knowledge that not everyone possesses, let alone more-than-humans or future generations. Some stakeholders may only have a voice hypothetically or by proxy, such as future generations and animals. Moreover, direct participation of every stakeholder who will suffer the consequences of the environmental impact of a specific AI system is impossible. Thus, this criterion requires democratic innovation, empathy and creativity.

Criterion (e) is a necessary addition to the above to avoid a tyranny of the majority (J. S. Mill, 1859). Costanza-Chock, for example, argues that a design that is beneficial for a majority may disadvantage minorities (Costanza-Chock, 2020). In the context of AI's environmental impact, pro-AI voices might come from privileged and thus more vocal groups, while marginalised and vulnerable actors will experience the greatest burdens, yet are often unheard in decision procedures. Therefore, disproportionate representation, giving more decision-making power to affected yet marginalised groups, may be necessary in the context of developing AI given its environmental impact (Young, 1990). Vulnerable groups that are most sensitive to the drawbacks of AI should have a significant, and perhaps decisive, influence on decisions about AI development.

*4.3 Justice as recognition*

In contrast to maldistribution, misrecognition refers to symbolic, cultural injustices (Álvarez & Coolsaet, 2020; Coolsaet & Néron, 2020). For example, misrecognition occurs in countries where gay marriage is illegal, even if a legal equivalent is available, because it implicitly devalues and judges same-sex relationships. Recognition of the equal value of romantic relationships, regardless of the sex of the partner(s) involved, is intrinsically important to people's identity, dignity and self-worth. As such, misrecognition, or cultural and symbolic injustices, deserves explicit attention as an analytically distinct tenet of justice. Moreover, misrecognition is crucial to consider, as it often underlies maldistribution or exclusion in decision-making procedures (Young, 1990).

AI systems are sociotechnical systems, as these technologies are the product of social (power) relations, norms, values, interests, priorities, and ideas (Bolte, 2023; Rohde et al., 2024). Thus, through AI, unjust relations, norms and value hierarchies can be reproduced or reinforced. The issue of misrecognition in AI is not new, as there is much research on the problem of bias in AI, reproducing -isms such as racism, ageism, sexism, and ableism, which are instances of misrecognition, as unjust social relations and norms are institutionalised and reproduced (e.g. Birhane, 2021; Waelen & Wieczorek, 2022). However, to our knowledge, no explicit links have yet been made between misrecognition and the environmental impact of AI.

In the context of the environmental impact of AI, it is crucial to reflect on the cultural values (or status order) and social relations (of recognition) that are institutionalised (i) in the institutions around AI development, such as the decision-making procedures and the design processes, and (ii) in the AI itself, that is, in the algorithm, the training data, and the output it generates. A tentative principle of justice for recognition in the context of the environmental impact of AI is:

*Developing an AI is equitable if the cultural values and social relations institutionalised in (i) the decision procedure and (ii) in the AI itself…*
- *(f) recognise all stakeholders as full partners in social life, and*
- *(g) contribute to an undistorted relation-to-self.*

Criterion (f) is based on Fraser's conception of recognition as status injury, or unjust institutionalised cultural values (Fraser, 2000; Fraser & Honneth, 2003). Cultural values are unjust, Fraser argues, if they prevent actors from participating equally in social life. She calls this criterion *participatory parity in social life*. Racism, for example, prevents people from being on an equal footing in social life, and so it is unjust to institutionalise racism. We consider it possible to stretch the notion of participatory parity in social life by including animals and ecosystems in our social life. Designs made for efficiency, monetary gain or human entertainment, at the expense of the environment, at least implicitly devalue non-humans and the natural environment (Moyano-Fernández & Rueda, 2024). In such cases, 'participatory parity in social life' can provide ammunition for critique.

First, AI can marginalise, ignore, disrespect, and undervalue actors through the decision-making procedures (i). Given the (predicted) environmental impact of an AI system, when decision-making about whether or not to develop the system fails to include voices of future generations, people of colour, people in the majority world, or more-than-human actors, it implicitly signals that these voices are irrelevant, do not matter, and are inferior, and that the impact on these actors is not important (enough) to consider (Holifield, 2012; Whyte, 2011). Unjust cultural norms can thus be built into the decision-making process.

Second, misrecognition can also be built into the design of AI itself (ii). Recently, Peter Singer and colleagues have argued that AI technologies often have species biases, as they implicitly devalue animals (for example in self-driving cars), misrepresent them (in image recognition), or reproduce speciest attitudes through, for example, algorithms recommending videos on animal cruelty (Hagendorff et al., 2023; Singer & Tse, 2023). Moreover, many design choices and choices related to whether to (not) develop an AI system with a certain environmental impact may implicitly devalue non-human animals, nature or vulnerable and underrepresented human actors (Moyano-Fernández & Rueda, 2024). For example, AI systems generally become increasingly accurate as they are trained longer on more data, which increases the environmental impact, leading to a trade-off between accuracy (and privacy, performance, and other benefits) and energy (E. Mill et al., 2022). This design choice entails, in effect, a trade-off between different values: the value of the benefits an AI brings versus the value of the environment and animals, future generations, Indigenous communities, and actors in vulnerable places and countries. Developed AI technologies and the regulations that enable them to embody these kinds of value hierarchies.

Criterion (g) formulates an alternative reason, besides participatory parity in social life, why misrecognition is unjust. Axel Honneth described that people are relational beings, and identity, autonomy and self-image are formed through the (mis)recognition of others (Honneth, 1995). Thus, one gains self-confidence, self-esteem, and self-respect when others treat and see us as valuable. When this is not the case, as in cases of racism or sexism, one can develop a distorted relation-to-self; one then lacks self-confidence, self-esteem or self-respect, which affects autonomy and well-being. Honneth argues that every human being needs and deserves an undistorted relation-to-self, and therefore, misrecognition is unjust. Institutionalised cultural values should contribute to a healthy relation-to-self, and this includes AI systems. This criterion provides another approach to assess institutionalised and often implicit social relations and values in (i) decision processes and (ii) in AI systems, and is included because it is complementary to criterion (f).

It makes little sense to speak of an undistorted relation-to-self in the context of animals or ecosystems. The relevance of Honneth's notion of misrecognition in AI mainly plays in the context of algorithmic bias, for example in face recognition technologies (Waelen, 2023). However, it is also relevant in the context of the environmental impact of AI, for three reasons. First, Honneth's theory of recognition also includes recognising the bodily integrity of people, and Schlosberg argued that there is potential in extending this argument to the need for recognition of the integrity of ecosystems and animals - harming the integrity of ecosystems and animals can also be considered misrecognition, and environmental injustice (Schlosberg, 2007). Second, one way of recognising the dignity and equal value of actors is through legislation, including by assigning rights, and avenues to recognise the rights of nature and ecosystems - and their dangers, limitations, and pitfalls - are being explored in both academia and beyond (Bellina, 2024). Third, justifications (or the lack of regulations) for AI systems rarely take into account the interests and value of, for example, indigenous groups and people in places vulnerable to climate change, which may result in them feeling disrespected and devalued.

## 5. Objections and responses

Before closing, we would like to anticipate three critiques on the proposed criteria that ought to inspire AI development guidelines, namely (1) the renewable energy critique, (2) the cause-effect critique, and (3) the inapplicability critique. In this section, we elaborate on these critiques and outline our response.

First, one may object that the environmental impact can be reduced significantly by using renewable energy sources, such as wind and solar energy, to power servers. As such, this would reduce the CO2 emissions generated by AI, decreasing the need for AI development guidelines to consider the environmental impact.

There are two objections to this critique. First, the environmental impact is broader than the CO2 emissions by energy consumption alone. The environmental impact includes water use, materials and resources for the necessary hardware, and the ensuing waste. Second, often, switching towards renewable energy sources does not always solve the problem. Datacentres may sign contracts with nearby wind and solar parks, and as a result, the renewable energy generated in those parks powers the datacentre instead of nearby local communities. In other words, before an all-renewable energy system is realised, data centres powered by renewables are an insufficient solution. Moreover, as more and more AI systems are developed and used, the demand for data centres and energy consumption increases, further fuelling the problem. Lastly, manufacturing wind and solar technologies are not without environmental impact either. As such, we argue that it is more sensible to critically reflect on each AI technology, whether it is truly beneficial and for which actors and geographies, instead of greenwashing AI technologies by pointing out that the data centre uses renewable resources.

Second, one might point out that it is almost impossible to calculate the precise environmental impact of a specific AI system. A specific natural disaster (such as a flood) can hardly be traced back causally to specific actions by specific actors (such as training a certain AI system). This is because the world is an incredibly complex system, and climate change is not linear but is connected to tipping points, in which small actions may set in motion a chain of disasters. As such, it cannot be predicted with certainty what the effects of developing a particular AI system will be. This causes problems for reflections on distributive justice and justice as recognition, as it is too uncertain what harm exactly will fall on which groups and regions. Because of this, it is not the responsibility of AI developers to engage in self-reflection but of governments to limit CO2 emissions in their jurisdiction.

However, we do not consider it necessary to be able to calculate the precise geographical effects of AI systems in order to reflect on environmental justice. In recent years, engineers and researchers have created tools and guides to calculate the environmental impact of AI systems, increasing transparency about the impact. The results may serve as an indication of AI's environmental impact. Moreover, it is generally known that the most vulnerable to the effects of CO2 emissions, climate change, mining raw materials, and waste disposal are located in the majority of the world. Also, it is usually possible to estimate who the users and beneficiaries of an AI system will be. We consider rough estimates about harms and benefits sufficient for critical reflection about whether AI should be developed or not, and if it will be developed, how to do so in ways that are environmentally just.

A third objection may be that the criteria are too vague. There are no hard (quantified) criteria for when there is a status order injustice or a distorted relation-to-self, for example. Moreover, it seems impossible and too contested to measure the benefits of AI in terms of joy, artistic value, and financial gain against harms such as CO2 emissions, given the incommensurability of values. As such, the criteria are insufficiently action-guiding and cannot be implemented in AI development guidelines.

We concede that the criteria are not yet quantified and remain vague to a certain extent. The aim of this paper was to provide a first iteration, and we encourage future research to further specify, discuss, and operationalise the criteria, towards implementation in ethics guidelines. However, as a first iteration of these principles, we argue that it is crucial to remain open towards different interpretations of environmental justice, and of different justifiable ways of making trade-offs between different values. We consider it a strength of the current iteration that the criteria are open for interpretation and dialogue, both within academia and society. Justice is a contested concept, and ethical guidelines should not be based on specific normative assumptions that are top-down translated into tick boxes but should be tools for ethical reflection and dialogue (Johnsson et al., 2014; Kiran et al., 2015). We want to stress, for example, the importance of a dialogue on the status of more-than-humans, the artistry of the output of DAL-E, and the value of ChatGPT for societies, given their environmental footprint. Our argument pertains to the importance of taking into account the environmental impact of AI in ethical assessments. Instead of abandoning this project because the criteria may seem vague, we encourage further research, both on developing the philosophical side and on translation towards AI development guidelines. We consider interpreting the criteria and guidelines, and what they imply for AI policy, as also a matter of political discussion and negotiation in democracies.

## 6. Conclusions

The environmental impact of AI cannot be underestimated. Besides the need for more research on the environmental impact of AI systems, we argue that it is crucial to take into account the environmental impact when ethically evaluating AI systems. In this contribution, we proposed expanding the scope of AI ethics towards considering the ethical implications of the environmental impact of AI, and including multispecies justice concerns, mitigating the anthropocentric bias in Western AI ethics.

Environmental justice literature articulates three categories of justice that refer to three elements that can be (un)just, namely distributions of benefits and environmental impacts, decision-making procedures, and institutionalised social norms and relations. These three elements formed the starting point for formulating conditions of the moral permissibility for developing AI systems, given its environmental impact. These conditions could guide the design of these technologies in ways that are environmentally just.

The conditions formulated in this article are a starting point for reflection, and they have implications for three groups of actors. First, AI developers and companies can assume their social responsibility by developing only those AI systems that are morally permissible and (more) environmentally just. This requires sincere critical reflection by companies on the distribution of burdens and benefits, the decision-making process, and implicit social norms and relations. Second, the biggest and most important role is for policymakers and regulators, as they can create guidelines or binding legislation for developing AI given its ecological impact and its implications for environmental justice. Finally, ethicists and political philosophers can contribute to the development of those rules and guidelines by conducting more research on the conditions of permissibility and justice that follow from increasing knowledge about the environmental impact of AI. This contribution takes a first step.